\documentclass[a4paper,11pt]{article}
\pdfoutput=1 % if your are submitting a pdflatex (i.e. if you have
\usepackage{jinstpub} % for details on the use of the package, please
\usepackage{float}
\newcommand{\figref}[1]{Fig.~\ref{#1}}

\title{\boldmath Search for neutrino non-standard interactions with ANTARES and KM3NeT-ORCA}

\author[a]{J. J. Hern\'{a}ndez Rey,}
\author[a,1]{N. R. Khan Chowdhury,\note{Corresponding author.}}
\author[a]{J. Manczak,}
\author[b]{S. Navas}
\author[a]{and J. D. Zornoza}

\affiliation[a]{IFIC - Instituto de Fisica Corpuscular (Univ. de Valencia - CSIC),\\ c/Catedr\'{a}tico Jos\'{e} Beltr\'{a}n, 2, E-46980 Paterna, Espa\~{n}a}
\affiliation[b]{University of Granada, Dpto. de F\'{i}sica Te\'{o}rica y del Cosmos,\\ Av. del Hospicio s/n, 18071 Granada, Espa\~{n}a}

\emailAdd{nafis.chowdhury@ific.uv.es}
\abstract{Non-standard interactions (NSIs) in the propagation of neutrinos in matter can lead to
significant deviations in neutrino oscillations expected within the standard 3-neutrino
framework. These additional interactions would result in an anomalous flux of neutrinos
observable at neutrino telescopes. The ANTARES detector and its next-generation successor,
KM3NeT, located in the abyss of the Mediterranean Sea, have the potential to measure
sub-dominant effects in neutrino oscillations, coming from non-standard neutrino interactions.
In this contribution, a likelihood-based search for NSIs with 10 years of atmospheric muon-neutrino
data recorded with ANTARES is reported and sensitivity projections for KM3NeT/ORCA, based on realistic detector simulations, are shown. The bounds obtained with ANTARES in the NSI $\mu - \tau$ sector constitute the most stringent limits up to date.}


\collaboration[c]{on behalf of the ANTARES and KM3NeT collaborations}

\proceeding{The 9$^{{th}}$ Very Large Volume Neutrino Telescope (VLV$\nu$T)\\
  18-21 of May, 2021\\
  Valencia (Virtual)}

\begin{document}
\maketitle
\flushbottom

\section{Neutrino propagation in presence of NSI}
When neutrinos propagate in matter, their evolution is affected by interactions with the medium, that result in the coherent forward elastic scattering of neutrinos. The overall effect can be described by effective potentials associated to the charged (CC) and neutral (NC) currents. In the case of neutrinos travelling through Earth, the only relevant potential is the one stemming from the electron neutrino components interacting with electrons in matter~\cite{Wolfenstein:1977ue}: $V_{CC} = \sqrt{2} \, G_{F} \, n_{e}$, where $G_{F}$ is the Fermi coupling constant and $n_{e}$ is the electron number density along the neutrino path. 

The presence of NSI in neutrino propagation can be described as an additional potential that will translate into an additional term in the neutrino propagation Hamiltonian:

\begin{equation*}
H^{NSI}= \frac{1}{2E_{\nu}} U M^2 U{^\dagger} + V_{CC} \, \text{diag(1,0,0)} + V_{CC} \, \frac{n_f}{n_e} \, {\epsilon},
\end{equation*}

\noindent where $U$, the PMNS mixing matrix, performs the rotation of the relevant mass matrix $M^2=\text{diag}(0, \Delta m^{2}_{21}, \Delta m^{2}_{31})$ in the neutrino flavour space. $n_f$ and $n_e$ are the fermion and electron number density along the neutrino path and neutrinos are assumed to interact with down quarks which are roughly three times as abundant as electrons, $n_{f} = n_{d} \approx$ 3 $n_{e}$. The matrix $\epsilon$ ($\epsilon_{\alpha \, \beta}$, $\alpha, \, \beta=$ e, $\mu$, $\tau$) gives the strength of the NSI. The diagonal terms of this matrix, if different from each other, can give rise to the violation of leptonic universality, while the off-diagonal terms can induce flavour-changing neutral currents, which are highly suppressed in the Standard Model (SM)~\cite{Liao:2016hsa}. 

\begin{figure}[ht]
%*** PRESENT FIGURE ***
\centering
\includegraphics[width=0.4\textwidth]{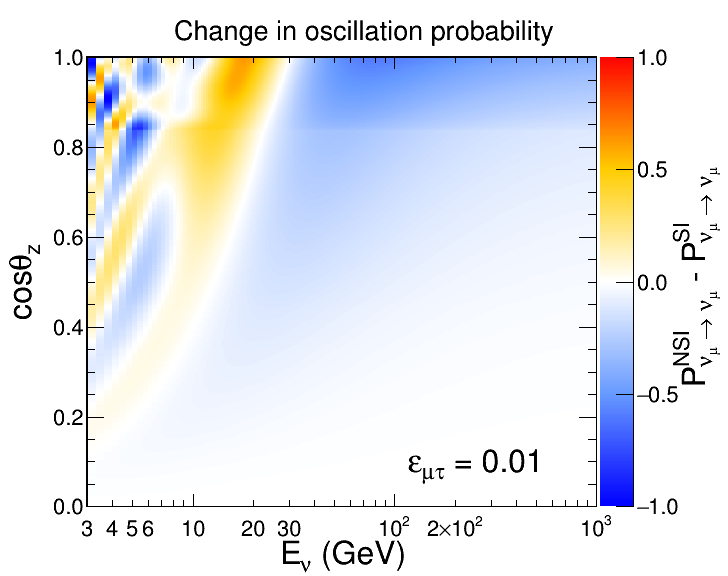}
\hspace{1cm}
\includegraphics[width=0.4\textwidth]{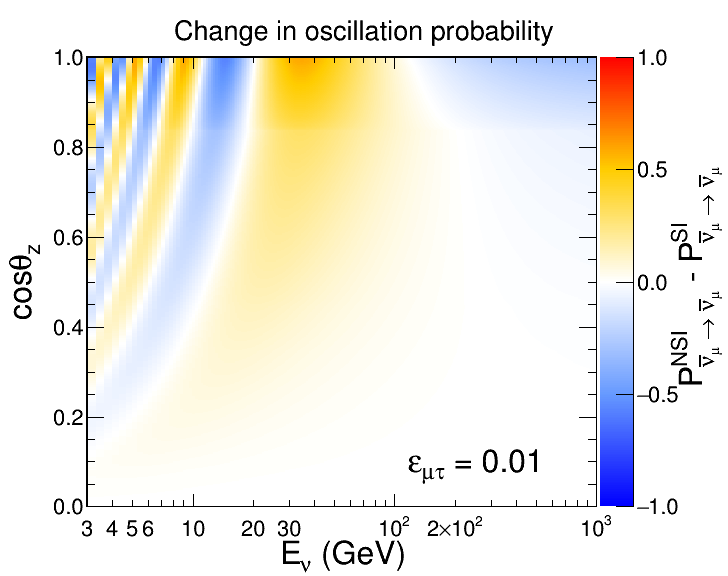}

\caption[]{NSI induced modifications in $\nu_{\mu}$ (left) and $\bar{\nu}_{\mu}$ (right) disappearance probabilities as a function of the neutrino energy and cosine of the zenith angle. %\footnotemark. 
The NSI test point has been set at $\varepsilon_{\mu\tau} = 0.01$. Normal Ordering (NO) of the neutrino masses is assumed.}
\label{fig:numuNSImutau}
\end{figure}

NSI are expected to have sub-dominant effects and therefore they would be observed as deviations from the distributions of the energy and arrival direction expected for standard oscillations.~\figref{fig:numuNSImutau} depicts the variation of the oscillation pattern in the muon disappearance channel due to NSIs. 

%\footnotetext{A value cos$\theta_{z} = 1$ corresponds to vertically up-going neutrinos traversing across the Earth core passing through all the density layers of the Earth, whereas cos$\theta_{z}$ = 0 corresponds to horizontally moving neutrinos.}

\section{The ANTARES and KM3NeT detector}
ANTARES~\cite{Zornoza:2012df} is a 0.01 km$^{3}$ deep-water water Cherenkov detector located at a depth of 2475 m in the Mediterranean Sea, 40 km offshore of Toulon (France). The basic detection component is a 17”-diameter pressure resistant sphere called Optical Module (OM), housing a 10” photomultiplier tube (PMT). Starting about 100 m from the
sea floor, the OMs are arranged in triplets vertically separated by about 15 m. The average horizontal separation between strings is $\sim$ 65 m and 12 such strings are anchored on the seabed to form a octagon with an effective mass of $\sim$10 Mtons.

%from Hammamatsu~\cite{Amram:2001mi} and associated electronics. Three OMs, staggered by $120^{\circ}$,  are grouped together, facing downwards at $45^{\circ}$ off axis, to form a storey. 25 vertical storeys compose a string. The vertical spacing between two adjacent storeys is $\sim$ 14.5 m and the first storey is at 100 m from the seabed. 

KM3NeT~\cite{Adrian-Martinez:2016fdl} is the next-generation upgrade of ANTARES, currently under construction in the Mediterranean Sea. 
Based on the granularity of the optical modules (to target different neutrino energy regimes), KM3NeT will house two detector at different locations: ORCA (for measuring neutrino properties) and ARCA (for doing neutrino astronomy). ORCA represents a three-dimensional array of $\sim$ 64,000 PMTs distributed among 115 detection strings  with 18 spherical Digital Optical Modules (DOMs) per line. Starting about 40 m from the
sea floor, the DUs of ORCA are 200 m high, horizontally separated by about 20 m, with 18 DOMs spaced 9 m apart in the vertical direction.

\section{Analysis}
 A first step is to compute the number of events corresponding to a specific oscillation hypothesis with or without NSI. The total number of charged current muon neutrino events expected at the detector for a certain runtime $t$ is given by:
\begin{equation*}
\begin{aligned}
    \frac{d^{2}N_{\mu}^{CC}}{dE \: d\cos\theta} =& \Big(\frac{d^{2}\phi_{\nu_{\mu}}}{dE \: d\cos\theta}P_{\mu\mu}+ \frac{d^{2}\phi_{\nu_{e}}}{dE \: d\cos\theta}P_{e\mu}\Big)\times \sigma^{CC}_{\nu_{\mu}} A^{CC}_{\nu_{\mu}} \times t \\
    &+ \Big(\frac{d^{2}\phi_{\bar{\nu}_{\mu}}}{dE \: d\cos\theta}P_{\bar{\mu}\bar{\mu}}+ \frac{d^{2}\phi_{\bar{\nu}_{e}}}{dE \: d\cos\theta}P_{\bar{e}\bar{\mu}}\Big)\times \sigma^{CC}_{\bar{\nu}_{\mu}}A^{CC}_{\bar{\nu}_{\mu}} \times t.
    \end{aligned}
\end{equation*}

$N_{\mu}$ is the number of detected muon events within the range of energy $dE$ and cosine of zenith angle $d\cos\theta$; $\phi_{\nu_{x}}$ ($\phi_{\bar{\nu}_{x}}$) is the atmospheric flux of neutrinos (antineutrinos) of flavour $x$ at the detector site; $P_{\alpha\beta}$ is the probability of oscillation of a neutrino flavour $\nu_{\alpha}$ to a neutrino flavour $\nu_{\beta}$; $\sigma^{CC}_{\nu_{\mu}}$ is the charged current cross-section of muon neutrino with nucleons in sea water; $A^{CC}_{\nu_{\mu}}$ is the energy and zenith angle dependent effective area of the detector corresponding to muon neutrinos undergoing CC interaction within the detector.

Depending on the Cherenkov signatures of the outgoing lepton from the $\nu_{e}-$, $\nu_{\mu}-$ and $\nu_{\tau}-$ CC and NC interactions, two distinct event topologies are observed at the detector: track-like and shower-like events. $\nu_{\mu}$ CC and $\nu_{\tau}$ CC interactions with muonic $\tau$ decays mostly account for the track-like topology. %, since the outgoing muon appears as a track within the detector. 
The shower-like topology corresponds to events from $\nu_{e}$ CC, $\nu_{\tau}$ CC interactions with non-muonic $\tau$ decays  and NC interactions of all flavours. The eight distributions indexed by interaction type $\mathcal{X}$ $\in \{(\nu_e,\, \overline{\nu}_e,\, \nu_{\mu},\, \overline{\nu}_{\mu},\, \nu_{\tau},\, \overline{\nu}_{\tau}$ - CC) and ($\nu,\, \overline{\nu}$ - NC)\} are merged and split into two distributions corresponding to the two event topologies: tracks and showers.

\begin{figure}
\begin{minipage}{.65\textwidth}
%\centering
\includegraphics[scale=0.28]{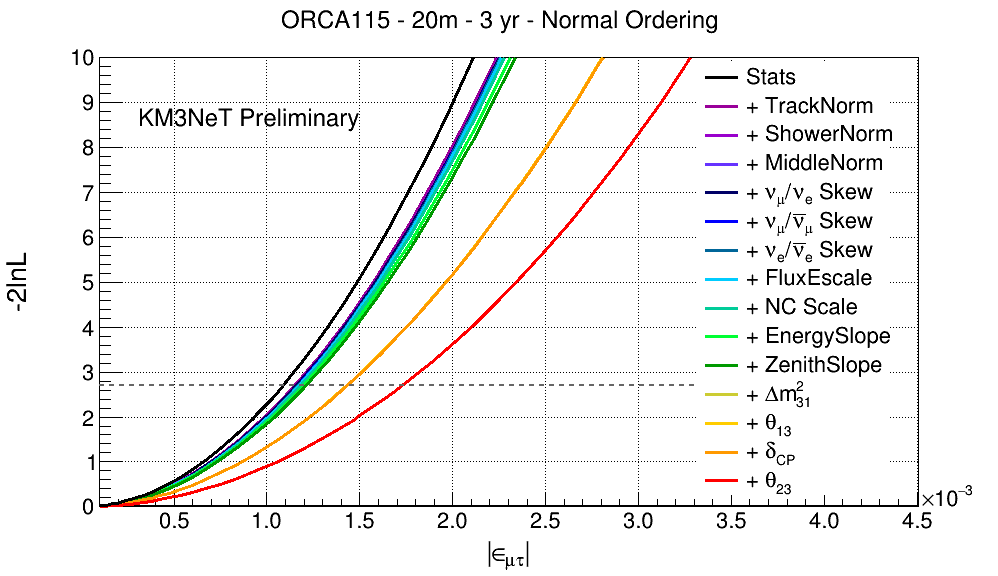}
\end{minipage}
\begin{minipage}{.34\textwidth}
   \caption{The list of systematics encountered in this analysis. Each colour coded curve corresponds to the effect of that particular systematic plus (\textbf{+}) the ones appearing on top of it being fitted simultaneously. The systematics are added incrementally in the sequence as they appear in the legends from violet to red. 
   } \label{table:systematicsorcasummary}
\end{minipage}
  \end{figure}

Assuming a Poissonian distribution of event numbers, the test statistic to estimate the sensitivity to a test parameter in the NSI model hypothesis is based on a binned likelihood approach~\cite{reid2003likelihood}:

\begin{equation*}
    -2LLR(\lambda, n) = 2\cdot\sum_{i\in \{bins\}} \bigg[\lambda_i(\bar{o}, \bar{s}) - n_i + n_i\ln\Big(\frac{n_i}{\lambda_i(\bar{o}, \bar{s})}\Big)\bigg] + \sum_{j \in \{syst\}}\frac{(s_{j} - \hat{s}_{j})^2}{2\sigma_{j}^2}.
    \label{eq:LLRsummary}
\end{equation*}

 where the number of predicted events $\lambda_i$ in the $i^{th}$ bin is a function of the set of oscillation parameters, $\bar{o}$, as well as on the the set of parameters related to systematic uncertainties, $\bar{s}$. The second term runs over penalty terms of the number, $j$, of nuisance parameters, $\hat{s}_{j}$ and $\sigma^{2}_{j}$ being the assumed prior and Gaussian standard deviation of the parameter $j$, respectively. The exhaustive list is shown in Fig.~\ref{table:systematicsorcasummary} and can be found in~\cite{KhanChowdhury:2021kce}.

\section{Results}
\begin{figure}
\begin{minipage}{.45\textwidth}
\centering
\includegraphics[width=\textwidth]{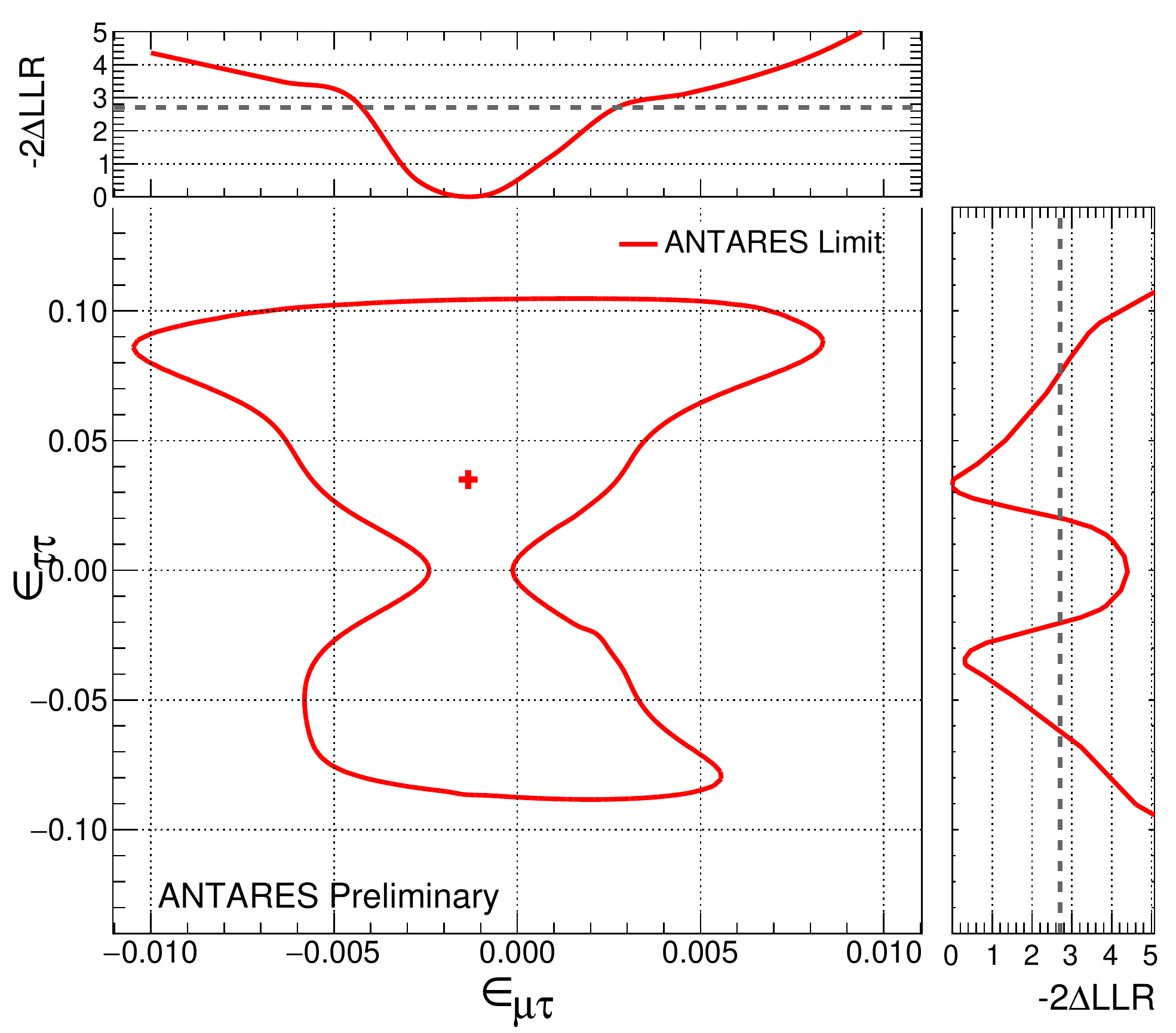}
\end{minipage}
\hspace{1cm}
\begin{minipage}{.45\textwidth}
\centering
\begin{table}[H]
\centering
% \newcolumntype{C}{>{\centering\arraybackslash}X}
%\resizebox{10cm}{3cm}{%

\begin{tabular}{|l c c |}

\hline
\hline
 \textbf{NSI }&\bf{NMO}&\textbf{ORCA (90\% C.L.)}  \\
\hline
\hline
$\epsilon_{e\mu}$ & NO & $(-1.7\times 10^{-2},\, 1.7\times 10^{-2})$ \\
                  & IO & $(-2.0\times 10^{-2},\, 2.0\times 10^{-2})$ \\

$\epsilon_{e\tau}$ & NO & $(-1.8\times 10^{-2},\, 2.1\times 10^{-2}$) \\
                   & IO & $(-3.1\times 10^{-2},\, 2.7\times 10^{-2}$) \\

$\epsilon_{\mu\tau}$ & NO & $(-1.7\times 10^{-3},\, 1.7\times 10^{-3}$) \\
                     & IO & $(-1.7\times 10^{-3},\, 1.7\times 10^{-3}$) \\

$\epsilon_{\tau\tau}$ & NO & $(-0.8\times 10^{-2},\, 1.1\times 10^{-2}$) \\
                      & IO & $(-1.1\times 10^{-2},\, 0.8\times 10^{-2}$)\\
\hline
\hline
\end{tabular}

\end{table}

\end{minipage}

\caption{\textbf{Left}: 90\% C.L. upper limits allowed after 10 years of ANTARES livetime obtained in this work are shown. The cross depicts the best-fit point obtained. \textbf{Right}: Bounds on NSI couplings at 90\% C.L. for two assumed mass orderings, for a runtime of 3 years of full ORCA comprising 115 DUs with 20 m horizontal DU spacing. Only one NSI parameter is considered at a time. }
\label{fig:limitsantaresallquad}
\end{figure}

 A total ANTARES livetime of 2830 days between years [2007, 2016] has been used~\cite{Albert:2018mnz, Ilenia}. 
The data was found consistent with standard oscillation hypothesis at $ 1.8\sigma$. \figref{fig:limitsantaresallquad} shows the 90\% C.L. upper limit obtained with 10 years of ANTARES. Limits on NSI matrix elements are extracted by profiling over the other variable:
\vspace{-0.5cm}

\begin{equation*}
   \begin{array}{rcl}
\centering
    -4.2 \times 10^{-3} < \epsilon_{\mu\tau} < 2.7 \times 10^{-3}  \qquad \qquad \qquad \qquad \quad &(\text{at}\, 90\% \, \textnormal{C.L}),\\
     -6.1\times 10^{-2} < \epsilon_{\tau\tau}  < -2.1 \times 10^{-2} \quad \text{and} \quad 2.1\times 10^{-2} < \epsilon_{\tau\tau}  < 7.3\times 10^{-2} \quad &(\text{at}\, 90\% \, \textnormal{C.L}).    
\end{array}
\end{equation*}

The limits for NSI obtained with 10 years of atmospheric muon disappearance data collected with ANTARES %is more stringent than allowed by current experimental limits, thereby 
constitutes the world's-best limits in the $\mu - \tau$ sector.

%For KM3NeT-ORCA, final event templates are simulated for a predicted exposure of 3~years of running. The expected event numbers are weighted according to various possible NSI signal hypotheses in order to explore the detection potential of ORCA towards NSIs. 
Expected bounds on NSIs after 3 years of running of full ORCA 115 DUs are collected in~\figref{fig:limitsantaresallquad}.
ORCA demonstrates an excellent potential to put tighter constraints on various NSI parameter spaces by one order of magnitude better than what is allowed by current experimental limits.

\section*{Acknowledgement}
We gratefully acknowledge the financial support of the Ministry of Science, Innovation and
Universities: State Program of Generation of Knowledge, ref. PGC2018-096663-B-C41 (MCIU /
FEDER), Spain.

%\newpage
\bibliographystyle{JHEP}
\bibliography{ref.bib}
\end{document}